\documentclass[aps,prb,twocolumn,preprintnumbers,amsmath,amssymb]{revtex4}%

\usepackage{graphicx}
\usepackage{dcolumn}
\usepackage{amsmath}
\usepackage{color}
\usepackage{hyperref}
\usepackage{academicons}

\newcommand{\RN}[1]{\textup{\uppercase\expandafter{\romannumeral#1}}}%

\newcommand{\orcid}[1]{\href{https://orcid.org/#1}{\textcolor[HTML]{A6CE39}{\aiOrcid}}}

\begin{document}

%\title{Use of light emitting diodes (LEDs) as an excitation source for fluorescence pressure indicators}
%
\title{Emergence of flat bands and their impact on superconductivity of Mo$_5$Si$_{3-x}$P$_x$}

\author{Rustem Khasanov$^1$}
 \email{rustem.khasanov@psi.ch}
\author{Bin-Bin Ruan$^2$}%[orcid=0000-0003-4642-7782]
 \email{bbruan@mail.ustc.edu.cn}
\author{Yun-Qing Shi$^{2,3}$}
\author{Gen-Fu Chen$^{2,3}$}
\author{Hubertus Luetkens$^1$}
\author{Zhi-An Ren$^{2,3}$}%[orcid=0000-0003-4308-7372]
\author{Zurab Guguchia$^1$}
 \affiliation{Laboratory for Muon Spin Spectroscopy, Paul Scherrer Institut, CH-5232 Villigen PSI, Switzerland}
 \affiliation{Institute of Physics and Beijing National Laboratory for Condensed Matter Physics, Chinese Academy of Sciences, Beijing 100190, China}
  \affiliation{School of Physical Sciences, University of Chinese Academy of Sciences, Beijing 100049, China}

\begin{abstract}
The first-principles calculations and measurements of the magnetic penetration depths, the upper critical field,  and the specific heat were performed for a family of Mo$_5$Si$_{3-x}$P$_x$ superconducotrs. First-principles calculations suggest the presence of a flat band dispersion, which gradually shifts to the Fermi level as a function of phosphorus doping $x$. The flat band approaches the Fermi level at $x\simeq 1.3$, thus separating Mo$_5$Si$_{3-x}$P$_x$ between the purely steep band and the steep band/flat band superconducting regimes. The emergence of flat bands lead to an abrupt change of nearly all the superconducting quantities. In particular, a strong reduction of the coherence length $\xi$ and enhancement of the penetration depth $\lambda$ result in nearly factor of three increase of the Ginzburg-Landau parameter $\kappa=\lambda/\xi$ (from $\kappa\simeq 25$ for $x\lesssim 1.2$ to $\kappa\simeq 70$ for $x\gtrsim 1.4$) thus initiating the transition of Mo$_5$Si$_{3-x}$P$_x$ from a moderate to an extreme type-II superconductivity.
\end{abstract}

%\pacs{74.70.Xa, 74.25.Bt, 74.45.+c, 76.75.+i}
\maketitle

%Introduction

Electrons with a narrow energy dispersion in the closed vicinity to the Fermi level are expected to demonstrate a broad variety of physics phenomena. Such electrons form a 'quasi-flat` bands, where the many-body effects dominate over the kinetic energy.
The famous example of a flat band physics is the the fractional quantum hole effect, where the Hall conductance of 2-dimensional  electrons shows precisely quantized plateaus at fractional values of $e^{2}/h$.
%($e$ is the elemental charge and $h$ is the Planck constant).
The quasiparticle excitations appears under a fractional filling of an electronic flat band which develops in the presence of a large magnetic field.\cite{Laughlin_PRB_1983, Goldman_Science_1995, Saminadayar_PRL_1997, Picciotto_Naturre_1997}
Another example is the twisted bilayer graphene, where flat bands are formed already at zero magnetic fields.\cite{Cao_Nature_2018} There, a flat electronic bands create large, many nanometre-size moir\'{e} unit cells, which folds and flattens the initial band structure of the material.\cite{Bistritzer_PNAS_2011}  Such flattening plays a crucial role in the physics of bilayer graphene and leads to appearance of a strong-coupling superconductivity, with the phase diagram resembling that of the high-temperature cuprates.\cite{Cao_Nature_2018,Li_Nature_2019}

Both above mentioned examples correspond to a rare case where the bands at the Fermi level stay nearly flat. In reality, the situation with the 'quasi-flat` band is more often to be realised. This is not so rare as might be thought, however. The recent careful search performed by Regnault {\it et al.},\cite{Regnault_Nature_2022} where more than 55 thousand compounds were analyzed, presented a catalogue of the naturally occurring three-dimensional stoichiometric materials with the quasi-flat bands around the Fermi level. It was found, in particular, that more than 5\% of all searched materials  host a flat band structures.

In relation to the superconducting materials, the importance of flat bands stems from substantial decrease of the  Fermi velocity, which may even tend to zero in a true flat band case.  Within the conventional Bardeen–Cooper–Schrieffer (BCS) approach this leads to a vanishingly small coherence length and the superfluid density, as well as to the extreme heavy and nearly immobile supercarriers. From the theory site, however, the emergence of flat bands stay in a favor to superconductivity by giving rise to a linear dependence of the transition temperature on the strength of the attractive interactions.\cite{Kopnin_PRB_2011, Ghanbari_PRB_2022, Shaginyan_EPL_2022, Tian_Nature_2023}  More interesting, the coexistence of flat and dispersive bands within the multi-band scenario lead to a strong enhancement of the transition temperature and  might potentially explain the phenomena of high-temperature superconductivity in cuprates and recently discovered hydride superconductors.\cite{Bussmann-Holder_CondMatt_2019}

In this work we probe the effect of band flattening on properties of Mo$_5$Si$_{3-x}$P$_x$ superconductors.
The emergence of flat bands leads to an abrupt change of nearly all superconducting quantities including: the transition temperature $T_{\rm c}$, the upper critical field $H_{\rm c2}$, the magnetic penetration depth $\lambda$, the coherence length $\xi$ and the superconducting energy gap $\Delta$.

\begin{figure*}[htb]
\includegraphics[width=0.9\linewidth]{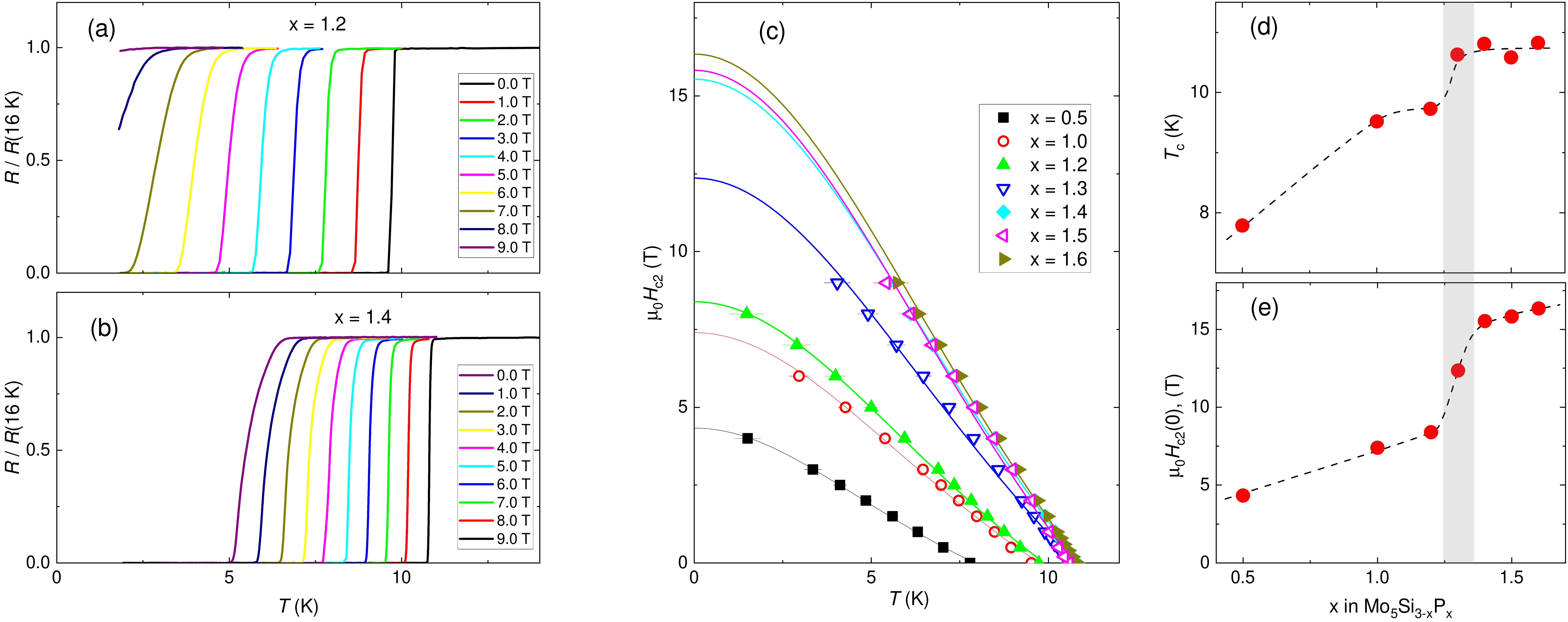}
\caption{ (a) and (b) The temperature dependencies of resistivity of Mo$_5$Si$_{3-x}$P$_{x}$ ($x=1.2$ and 1.4) samples measured in magnetic fields ranging from 0.0 to 9.0~T. (c) Temperature dependencies of the upper critical field $H_{\rm c2}$. The lines are fits of Eq.~\ref{eq:GL_Hc2} to the data. (d) Doping dependence of the transition temperature $T_{\rm c}$. (e) Doping dependence of the zero-temperature value of the upper critical field $H_{\rm c2}(0)$. The lines in (d) and (e) are guides for the eye. The grey stripe represents the region of an abrupt change of $T_{\rm c}$ and $H_{\rm c2}(0)$.}
 \label{fig:Resisitivity}
\end{figure*}

%Experimental

Polycrystalline Mo$_5$Si$_{3-x}$P$_x$ samples with $x=0.5$, 1.0, 1.2, 1.3, 1.4, 1.5, and 1.6 were prepared by a solid-state reaction method.\cite{Ruan_SciChinaMater_2022}
%The crystal structure and phase purity were checked by powder x-ray diffraction, confirming the tetragonal structure (space group $I4/mcm$).
In addition to the main Mo$_5$Si$_{3-x}$P$_x$ phase, a small amount of impurity phase Mo$_3$P (from $\simeq 5$ to 10\%) was detected.\cite{Supplemental_Part}
The first-principles calculations were performed based on the density functional theory, as implemented in the Quantum ESPRESSO package.\cite{Quantum-Espresso}
The superconducting properties were characterized by electrical resistivity and heat capacity measurements, performed on a Quantum Design PPMS (physical property measurement system).
The muon-spin rotation/relaxation ($\mu$SR) measurements were carried out at the $\pi$M3 beam line by using GPS (General Purose Surface) $\mu$SR spectrometer (Paul Scherrer Institut, Villigen, Switzerland).\cite{Amato_RSI_2017}
In this study, mostly the transverse-field (TF) $\mu$SR measurements were performed, which allowed to determine the temperature evolution of the magnetic penetration depth. %All the TF-$\mu$SR spectra were collected in the field cooling mode upon heating and
The $\mu$SR data were analyzed by means of the Musrfit software package.\cite{MUSRFIT}
The details of the sample preparation, the first-principles calculations, as well as the x-ray, resistivity, specific heat, and $\mu$SR experiments are provided in the Supplemental part.\cite{Supplemental_Part}

%Results

The doping evolution of the upper critical field $H_{\rm c2}$  was studied by means of resistivity. Figures~\ref{fig:Resisitivity}~(a) and (b) show the resistivity curves normalized to the values at $T=16$~K [$R(T)/R(16{\rm ~K})$] measured in magnetic fields ranging from 0.0 to 9.0~T for two representative phosphorus dopings $x=1.2$ and $x=1.4$, respectively. Obviously, the two closely doped samples, which have nearly similar superconducting transition temperatures at zero applied field ($\mu_0H_{\rm ap}=0.0$~T), reacts differently on the magnetic field. As an example, at $\mu_0H_{\rm ap}=9.0$~T the $x=1.2$ sample stays in a normal state down to $T\simeq 1.75$~K, while the $x=1.4$ one superconducts below $\simeq 5$~K.

The upper critical field values defined from $R(T)$ measurement curves are summarized in Fig.~\ref{fig:Resisitivity}~(c). Here, $T_{\rm c}(H_{\rm ap})$'s were determined from the mid point of $R(T,H_{\rm ap})$ curves [{\it i.e.} as the value where $R(T)/R(16{}\rm ~K)=0.5$]. The solid lines correspond to the fits of the Ginzburg-Landau model
\begin{equation}
  H_{\rm c2}(T) = H_{\rm c2}(0)\ \frac{1-(T/T_{\rm c})^2}{1+(T/T_{\rm c})^2}
  \label{eq:GL_Hc2}
\end{equation}
to the experimental $H_{\rm c2}(T)$ data.
The values of $T_{\rm c}$'s at $H_{\rm ap}=0$ and $H_{\rm c2}(0)$'s from fits of Eq.~\ref{eq:GL_Hc2} are presented in Figs.~\ref{fig:Resisitivity}~(d) and (e) as a function of phosphorus doping $x$. An abrupt change of both parameters at $x\simeq 1.3$ is clearly visible.

\begin{figure*}[htb]
\includegraphics[width=0.9\linewidth]{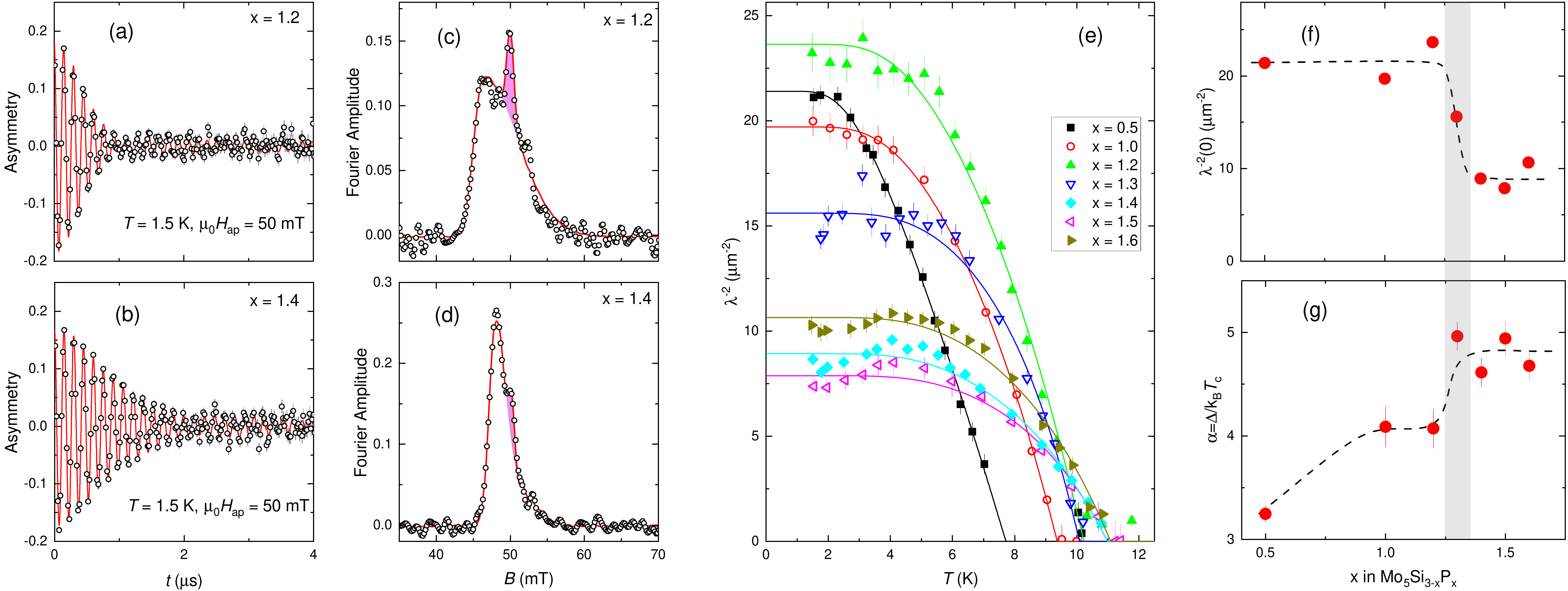}
\caption{ (a) and (b) The muon-time spectra of Mo$_5$Si$_{3-x}$P$_{x}$ ($x=1.2$ and 1.4) samples measured at $T=1.5$~K and $\mu_0H_{\rm ap}=50$~mT. The solid lines are fits of Eq.~\ref{eq:asymmetry} to the data. (c) and (d) The Fourier transforms of asymmetry spectra presented in panels (a) and (b). Red peaks denote the background contribution  originating from muons missing the sample. (e) Temperature evolution of $\lambda^{-2}$. The solid lines are fits of Eq.~\ref{eq:lambda-s} to the data. (f) Doping dependence of $\lambda^{-2}(0)$. (g) Doping dependence of the gap to $T_{c}$ ratio $\alpha=\Delta/k_{\rm B} T_{\rm c}$. The lines in (f) and (g) are guides for the eye. The grey stripe represents the region of an abrupt change of $\lambda^{-2}(0)$ and $\alpha$. }
 \label{fig:mSR}
\end{figure*}

The temperature dependencies of the magnetic penetration depth $\lambda$ were studied in TF-$\mu$SR experiments. Measurements were performed in the field cooling mode at the applied field $\mu_0H_{\rm ap}=50$~mT.  Representative TF-$\mu$SR time-spectra of $x=1.2$ and 1.4 samples at $T\simeq 1.5$~K ({\it i.e.} below $T_{\rm c}$) are shown in Figs.~\ref{fig:mSR}~(a) and (b). A strong damping reflects the inhomogeneous field distribution $P(B)$ caused by the formation of the flux-line lattice (FLL). The broad asymmetric  distribution is clearly visible in Figs.~\ref{fig:mSR}~(c) and (d), where the Fourier transform  of the corresponding TF-$\mu$SR data are shown. The $P(B)$ distributions demonstrate all characteristic features of a well arranged FLL, namely the cut-off at low-field, the extended tail to the higher field values and shift of the $P(B)$ peak below $\mu_0H_{\rm ap}$.\cite{Maisuradze_JPCM_2008} Note that the narrow peak at the applied field position ($B_{\rm ap}=\mu_0H_{\rm ap}$) originates from muons missing the sample.

The width of $P(B)$  within the FLL in the limit of $H_{\rm ap}\ll H_{\rm c2}$ [as is the case for our studies, see Figs.~\ref{fig:Resisitivity}~(c) and (e)] is primarily determined by the value of the magnetic penetration depth $\lambda$.\cite{Maisuradze_JPCM_2008, Brandt_PRB_1988, Brandt_PRB_2003} Comparison of $P(B)$'s presented in Figs.~\ref{fig:mSR}~(c) and (d) suggests that $x=1.2$ sample has stronger broadening ({\it i.e.} smaller $\lambda$ value) compared to that of $x=1.4$ one.
The analysis of TF-$\mu$SR data was performed by considering the presence of a main Mo$_5$Si$_{3-x}$P$_{x}$ phase (denoted as 's') and two background contributions ('bg,1' and 'bg,2'), respectively. The 'bg,1' contribution originates from the impurity Mo$_3$P phase (which superconducts at $T_{\rm c}\simeq 5.5$~K, Refs.~\onlinecite{Matthias_PR_1954, Blaugher_JPCS_1965, Shang_PRB_2019}), while the 'bg,2' one is caused by muons missing the sample ({\it i.e.} stopped at the sample holder and the cryostat windows). The following functional form was used:
\begin{eqnarray}
  A(t)& = & A_{\rm s}\; {\rm SkG}(t,B_{\rm s},\sigma_+,\sigma_-)+A_{\rm bg,1}\; e ^{-\sigma_{\rm bg,1}^2t^2/2}  \\
  && \times\cos(\gamma_\mu B_{\rm bg,1}+\phi)+A_{\rm bg,2} \cos(\gamma_\mu B_{\rm ap}+\phi). \nonumber
  \label{eq:asymmetry}
\end{eqnarray}
Here  $A_{s}$ ($\sim 90$\%), $A_{\rm bg,1}$ ($\sim 10$\%), and $A_{\rm bg,2}$ ($\sim 1$\%) are the initial asymmetries, and $B_{\rm s}$, $B_{\rm bg,1}$, and $B_{\rm ap}$ are the internal fields of each particular component. $\gamma_\mu=2\pi\cdot 135.53$~MHz/T is the muon gyromagnetic ratio, $\phi$ is the initial phase of the muon-spin ensemble, and $\sigma$ is the Gaussian relaxation rate. The sample contribution was fitted with the Skewed Gaussian  function [SkG($B,\sigma_+,\sigma_-$)], which accounts for the asymmetric $P(B)$ distribution within the FLL.\cite{Suter_SKG, Khasanov_FeSe-Intercalated_PRB_2016} The second central moment of the sample contribution $\langle \Delta B^2\rangle_{\rm s}$ was obtained from the fitted $\sigma_+$ and $\sigma_-$ values.\cite{Suter_SKG}  The inverse squared magnetic penetration depth was further calculated as $\lambda^{-2}[\mu{\rm m}^{-2}]=9.32\times \sqrt{\langle \Delta B^2\rangle_{\rm s} -\sigma_{\rm nm}^2} \; [\mu{\rm s}^{-1}]$. \cite{Khasanov_FeSe-Intercalated_PRB_2016, Brandt_PRB_2003} Here $\sigma_{\rm nm}$ is the nuclear moment contribution which is determined from the measurements above $T_{\rm c}$.\cite{Maisuradze_JPCM_2008, Khasanov_FeSe-Intercalated_PRB_2016}

The temperature dependencies of the inverse squared magnetic penetration depth of Mo$_5$Si$_{3-x}$P$_x$ samples are presented in Fig.~\ref{fig:mSR}~(e). For all doings $\lambda^{-2}(T)$'s demonstrate saturation for $T\lesssim4$~K ({\it i.e.} for temperature below $\sim$ 1/3 of $T_{\rm c}$), which is consistent with the formation of a fully gapped state.
The solid lines represent best fits within the $s-$wave BCS model:\cite{Tinkham_book_1975}
\begin{equation}
\frac{\lambda^{-2}(T)}{\lambda^{-2}(0)} =  1
+2\int_{\Delta(T,\varphi)}^{\infty}\left(\frac{\partial
f}{\partial E}\right)\frac{E\
dEd\varphi}{\sqrt{E^2-\Delta(T)^2}}~.
 \label{eq:lambda-s}
\end{equation}
Here $f=[1+\exp(E/k_BT)]^{-1}$ is  the Fermi function and $\Delta(T)=\Delta(0) \tanh\{1.82[1.018(T_{\rm c}/T-1)]^{0.51}\}$ is the temperature dependent superconducting gap.\cite{Khasanov_PRL_La214_07} $\lambda^{-2}(0)$ and $\Delta(0)$ are the zero-temperature values of the inverse squared penetration depth and the superconducting gap, respectively. The dependencies of the fit parameters, namely $\lambda^{-2}(0)$ and $\alpha=\Delta(0)/k_{\rm B}T_{\rm c}$ on the phosphorus content $x$ are summarized in Figs.~\ref{fig:mSR}~(f) and (g), respectively. A step-like change of both parameters at $x\simeq 1.3$ is clearly visible.

The results obtained in resistivity (Fig.~\ref{fig:Resisitivity}) and TF-$\mu$SR (Fig.~\ref{fig:mSR}) experiments imply that the major superconducting quantities, namely the transition temperature $T_{\rm c}$, the upper critical field $H_{\rm c2}$, the magnetic penetration depth $\lambda$, and the energy gap $\Delta(0)$ demonstrate an abrupt change at $x\simeq 1.3$. Two possible scenario can be considered. The first one assumes a formation of a competing ordered state, where part of the carries are gapped due to competing interactions and, therefore, become unaccessible  for the Cooper pair formation. As an example of such states, one may refer to the charge-density wave (CDW) or spin-density wave (SDW) type of orders, which are widely detected for cuprate, kagome, and Fe-based superconducting families.\cite{Fradkin_RMP_2015, Wu_Nature_2011, Fernandes_NatPhys_2014, Dai_RMP_2015, Neupert_NatPhys_2022, Gupta_CommPhys_2022, Mielke_Nature_2022, Khasanov_PRR_2022, Guguchia_NatCom_2023}
These scenario is not plausible here, since: (i) The specific heat experiments reveal the absence of an abrupt change of the density of states at the Fermi level [$N(E_{\rm F})$] in the vicinity of $x\simeq 1.3$.\cite{Supplemental_Part} (ii) The zero-field $\mu$SR experiments do not detect any kind of magnetism,\cite{Supplemental_Part} thus implying that the SDW type of order does not come into play.
(iii) The resistivity experiments reported in Ref.~\onlinecite{Ruan_SciChinaMater_2022} do not detect any features at the  normal-state resitivity curves up to $T\simeq 300$~K.

\begin{figure}[htb]
\includegraphics[width=1.0\linewidth]{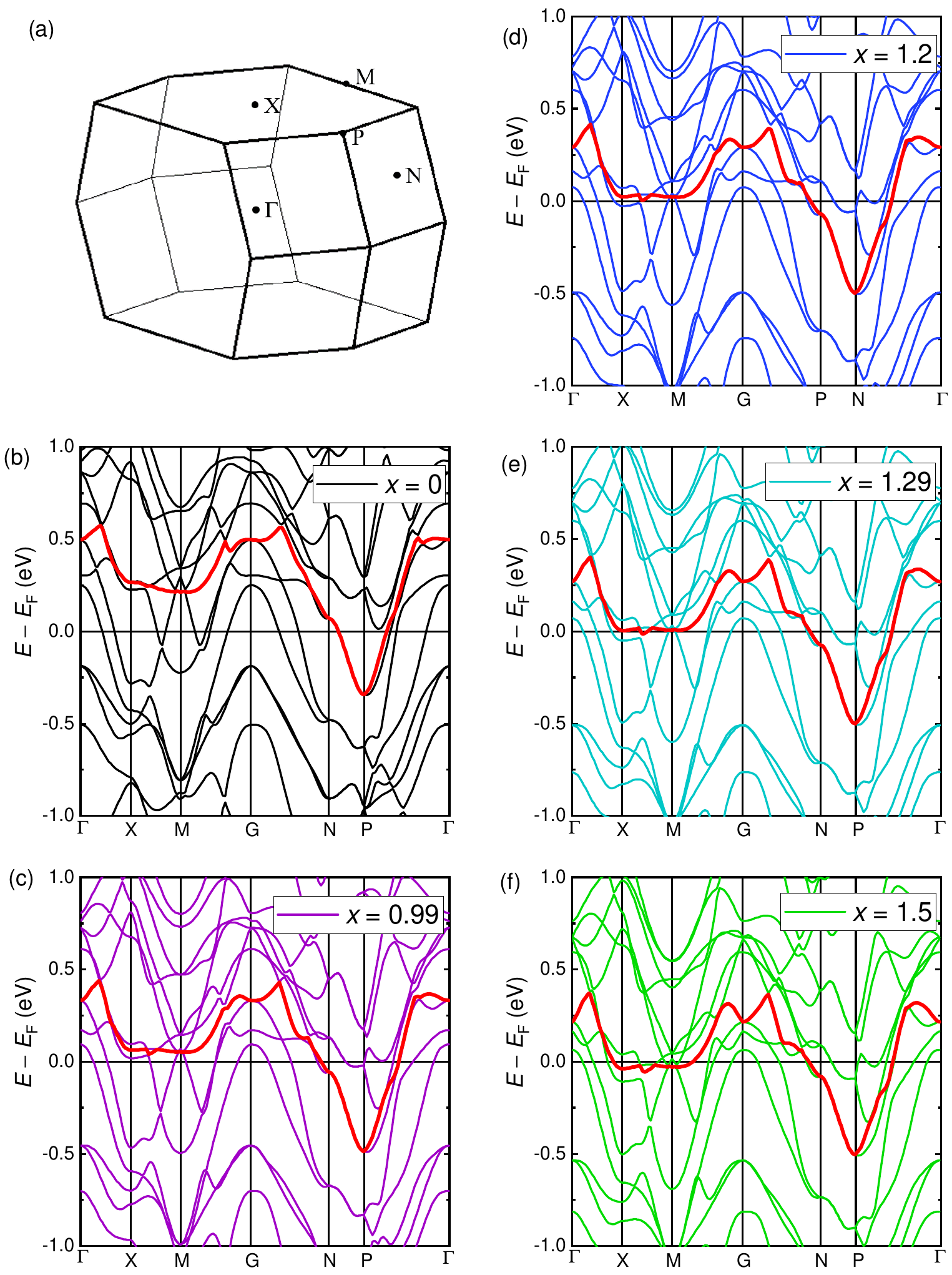}
\caption{ (a) Brillouin zone of Mo$_5$Si$_{3-x}$P$_x$. The high-symmetry points are labeled as $\Gamma$, $X$, $M$, $N$, and $P$. (b)-(f) The electronic band structures of Mo$_5$Si$_{3-x}$P$_x$ with $x=0.0$, 0.99, 1.2, 1.29, and 1.5. The band with extended flattened portion is denoted by the red color. }
 \label{fig:DFT}
\end{figure}

The second scenario assumes an emergence of flat bands at the Fermi level. In order to demonstrate this, the set of band structure calculations of Mo$_5$Si$_{3-x}$P$_x$, with a special attention paid to the critical doping region at $x\simeq 1.3$, was performed. The results are presented in Fig.~\ref{fig:DFT}. The shape of the first Brillouin zone and positions of the high symmetry points ($\Gamma$, $X$, $M$, $N$, and $P$) are shown in Fig.~\ref{fig:DFT}~(a). The  major feature of the electronic structure is the band denoted by the red color, which has a substantial flattened portion. At zero doping, the flat dispersion sets at energy $\simeq0.25$~eV above the Fermi energy ($E_{\rm F}$), Fig.~\ref{fig:DFT}~(b). With increasing the phosphorus content $x$, $E_{\rm F}$ shifts to the higher energies and at $x\simeq 1.3$ the flat band dispersion approaches the Fermi level, Fig.~\ref{fig:DFT}~(e). This clearly indicates that the transition from the purely steep band to the steep band/flat band scenario is {\it realised} in Mo$_5$Si$_{3-x}$P$_x$.
It is also remarkable that the doping level, at which the flat band approaches $E_{\rm F}$, coincides with that, where the abrupt changes of various superconducing quantities take place [Figs.~\ref{fig:Resisitivity}~(d), \ref{fig:Resisitivity}~(e), \ref{fig:mSR}~(f), and \ref{fig:mSR}(g)].

The effects of band flattening on the two fundamental superconducting length scales, namely the magnetic penetration depth $\lambda$ (which defines a distance for magnetic field decay) and coherence length $\xi$ (which determines dimensions of the Cooper pairs), might be understand in relation with the corresponding changes of the Fermi velocity $v_{\rm F}$. Note that all these quantities are obtainable from the above presented data: the value of $\xi$ could be calculated from the measured $H_{\rm c2}$ by using the Ginzburg-Landau expression $H_{\rm c2}=\Phi_0/2\pi\xi^2$, \cite{Kittel_book_2013} $\lambda$ is measured directly in TF-$\mu$SR experiments, while the Fermi velocities might be estimated from the electronic structure as the first derivative of the band dispersions at $E_{\rm F}$.
%
%The band flattening lead do decrease of the Fermi velocity which is proportional to the first derivative of the band dispersion in the vicinity of $E_{\rm F}$.

For a single-band superconductor and within the conventional BCS scenario the zero-temperature values of the coherence length and the penetration depth follow the well-known relations:
\begin{equation}
\xi(0)=\frac{\hbar \langle v_{\rm F}\rangle}{\pi \Delta(0)} = \frac{1}{\pi \alpha}\frac{\hbar \langle v_{\rm F}\rangle}{k_{\rm B}T_{\rm c}}
 \label{eq:xi0}
\end{equation}
and
\begin{equation}
\lambda(0)=\sqrt{\frac{m^\ast}{\mu_0 n_s e^2}} = \sqrt{\frac{\hbar \langle k_{\rm F}\rangle}{\langle v_{\rm F}\rangle n_s e^2}}
\label{eq:lambda0}
\end{equation}
Here $\hbar$ is the reduced Planck constant, $\langle v_{\rm F}\rangle$ is the average value of the Fermi velocity, $n_{\rm s}$ is the charge carrier concentration, $m^\ast=\hbar \langle k_{\rm F}\rangle/\langle v_{\rm F}\rangle$ is the effective carrier mass and $\langle k_{\rm F}\rangle$ is the averaged Fermi wave vector.
\begin{figure}[htb]
\includegraphics[width=0.75\linewidth]{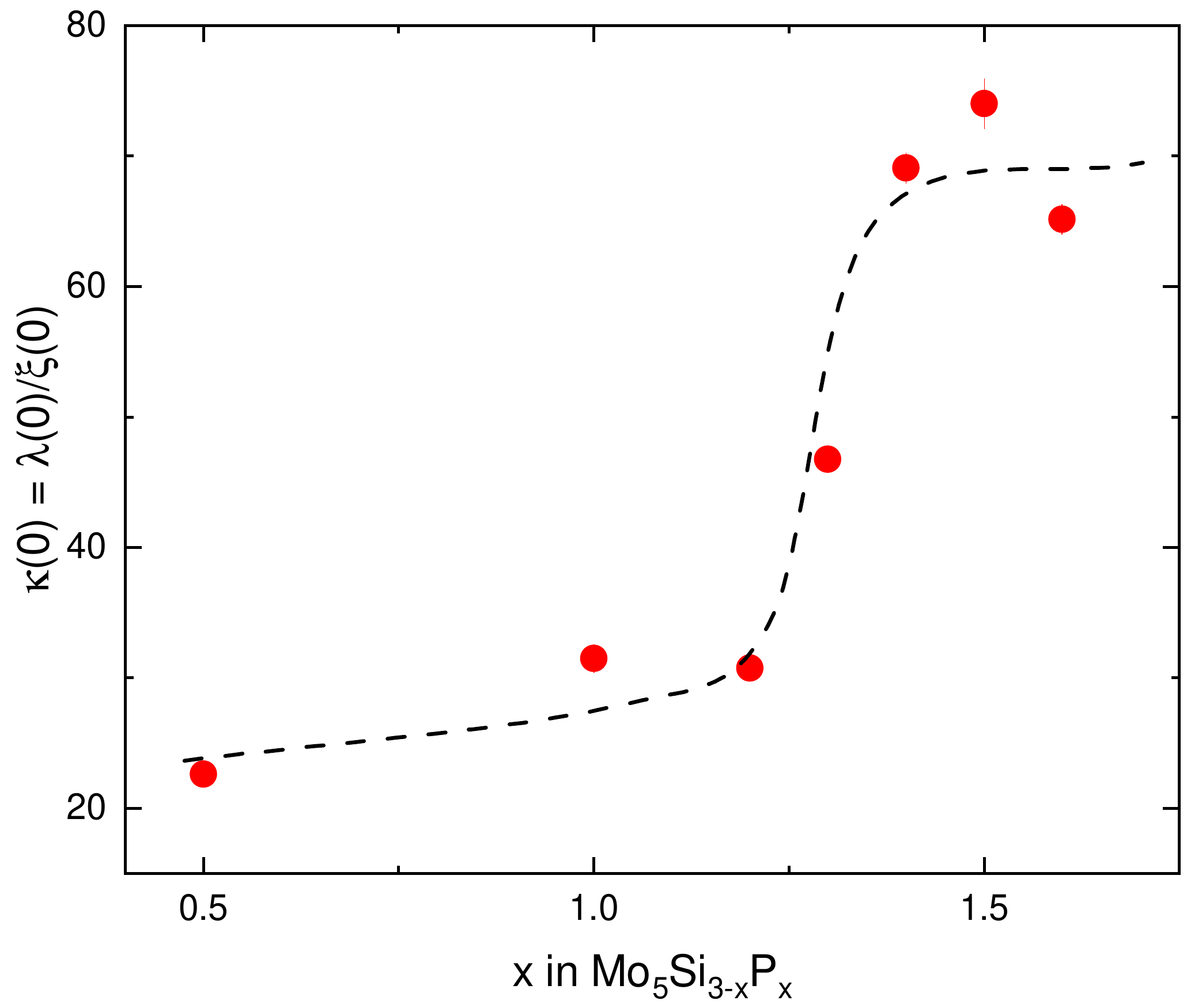}
\caption{ The dependence of the Ginzburg-Landau parameter $\kappa=\lambda(0)/\xi(0)$ of Mo$_5$Si$_{3-x}$P$_x$ on the phosphorus content $x$.}
 \label{fig:kappa}
\end{figure}
By having only limited validity for Mo$_5$Si$_{3-x}$P$_x$, which is definitively not a single-band, but a multi-band superconductor [see Fig.~\ref{fig:DFT}], the above equations allow to capture the main features of our experimental observation:
(i) The emergence of flat bands at the Fermi level, occurring  at $x\gtrsim 1.3$ [Fig.~\ref{fig:DFT}~(d)], leads to a sudden decrease of $\langle v_{\rm F}\rangle$. Following Eqs.~\ref{eq:xi0} and \ref{eq:lambda0}, this requires the coherence length $\xi(0)$ to increase, and $\lambda(0)$ to decrease accordingly. The corresponding effects on the measured quantities $H_{\rm c2}(0)$  and $\lambda^{-2}(0)$  are just opposed to that of $\xi(0)$ and $\lambda(0)$, in agreement with the experimental observations [Figs.~\ref{fig:Resisitivity}~(e) and \ref{fig:mSR}~(f)].
(ii) An opposed band flattening effect on $\xi(0)$ and $\lambda(0)$ would imply a strong change of the Ginzburg-Landau parameter $\kappa=\lambda/\xi$. This is demonstrated in Fig.~\ref{fig:kappa}, where $\kappa(0)$ increases by nearly three times from $\simeq 25$ for $x\leq1.2$ to $\simeq 70$ for $x\geq 1.4$.

To conclude, the first-principles calculations and the measurements of the magnetic penetration depths, the upper critical field and the specific heat for newly discovered superconductor family Mo$_5$Si$_{3-x}$P$_x$ were conducted. The emergence of flat bands for the phosphorus content $x$ exceeding $\simeq1.3$ leads to an abrupt change of nearly all superconducting quantities. In particular, the transition temperature $T_{\rm c}$ increases by $\simeq 15$\%: from $\simeq~9.5$ to nearly 11~K, the upper critical field $H_{\rm c2}$  is more than doubles: from $\simeq 7$ to $\simeq 16$~T, while the inverse squared penetration depth (which is normally  considered to be a measure of the supercarrier concentration) decreases by more than twice: from $\simeq 22$ to $\simeq 9$~$\mu$m$^{-2}$.
Our results suggests that the Mo$_5$Si$_{3-x}$P$_x$ superconducting family become a unique system with easily accessible interplay between the steep band and the steep band/flat band physics. The high transition temperatures and moderate critical fields makes this system ideal for understanding the role played by band flattening in the superconducting mechanism.

Z.G. acknowledges support from the Swiss National Science Foundation (SNSF) through SNSF Starting Grant (No. TMSGI2${\_}$211750).
Z.R. acknowledges supports from the National Key Research and Development Program of China (Grant Nos. 2018YFA0704200 and 2021YFA1401800) and the National Natural Science Foundation of China (Grant No. 12074414).

\end{document}